\documentclass[preprint,aps,floatfix,nofootinbib,superscriptaddress]{revtex4-1}
\pdfoutput=1

\usepackage{amsmath, amssymb, amsfonts, amsthm, latexsym, epsfig, mathrsfs, xcolor, bbm}

\usepackage[inline]{enumitem}

\usepackage{setspace}
\usepackage[marginal, multiple]{footmisc}

\usepackage[T1]{fontenc}
\usepackage[utf8]{inputenc}
\usepackage{lmodern}

\usepackage[colorlinks, allcolors=blue!70!black, linktocpage]{hyperref}

\setlist[enumerate]{noitemsep, label=(\arabic*), ref=(\arabic*)}

\numberwithin{equation}{section}

\usepackage{cleveref}

\usepackage{microtype}

\crefname{equation}{Eq.}{Eqs.}
\creflabelformat{equation}{(#2#1#3)}

\crefname{section}{Sec.}{Sec.}
\crefname{appendix}{Appendix}{Appendices}
\crefname{figure}{Fig.}{Figs.}

\crefname{definition}{Def.}{Defs.}
\crefname{prop}{Prop.}{Props.}
\crefname{lemma}{Lemma}{Lemmas}
\crefname{corollary}{Cor.}{Cors.}
\crefname{thm}{Theorem}{Theorems}
\crefname{remark}{Remark}{Remarks}

\makeatletter
\def\p@subsection{}
\def\p@subsubsection{}
\makeatother

\let\OLDtableofcontents\tableofcontents
\renewcommand\tableofcontents[1]{%
    {\baselineskip 0.5ex %
	\OLDtableofcontents{#1}}%
}

\let\OLDfootnote\footnote
\renewcommand\footnote[1]{%
        \setlength{\footnotesep}{0.55\baselineskip}%
        {\footnotesize \OLDfootnote{#1}}%
}

\let\OLDthebibliography\thebibliography
\renewcommand\thebibliography[1]{%
        \setstretch{1.079} % 1/2 goldenratio
        \OLDthebibliography{#1}%
        \small %
        \setlength{\itemsep}{0.2\baselineskip} % goldenratio for separating items %
}

\newcommand{\scri}{\mathscr{I}}

\newcommand{\lie}{\pounds}

\newcommand{\ms}{\mathscr}
\newcommand{\mf}{\mathfrak}
\newcommand{\bb}{\mathbb}

\newcommand{\df}[1]{\boldsymbol{#1}}

\newcommand{\eqsp}{\, ,\quad} %shrtct for space in eqns

\newcommand{\lb}{\left}
\newcommand{\rb}{\right}

\let\oldlim\lim
\renewcommand{\lim}{\oldlim\limits}

\renewcommand{\bar}{\overline}

\newcommand{\lra}{\leftrightarrow}

\def\hhmm{\number\hh:\ifnum\mm<10{}0\fi\number\mm}

\def\be{\begin{equation}}
\def\ee{\end{equation}}

\begin{document}

\setstretch{1.2}

\title{Extensions of the asymptotic symmetry algebra of general relativity}

\author{\'Eanna \'E. Flanagan}
\email{eef3@cornell.edu}
\affiliation{Department of Physics, Cornell University, Ithaca, New York, 14853, USA}

\author{Kartik Prabhu}
\email{kartikprabhu@cornell.edu}
\affiliation{Cornell Laboratory for Accelerator-based Sciences and Education (CLASSE),Cornell University, Ithaca, New York, 14853, USA}

\author{Ibrahim Shehzad}
\email{is354@cornell.edu}
\affiliation{Department of Physics, Cornell University, Ithaca, New York, 14853, USA}

%\date{{\color{magenta}{Draft updated by IS \today{}; printed \today{}}}}

\begin{abstract}
We consider a recently proposed extension of the Bondi-Metzner-Sachs
algebra to include arbitrary infinitesimal diffeomorphisms on a
\(2\)-sphere. To realize this extended algebra as asymptotic
symmetries, we work with an extended class of spacetimes in which the
unphysical metric at null infinity is not universal. We show that the
symplectic current evaluated on these extended symmetries is divergent
in the limit to null infinity. We also show that this divergence
cannot be removed by a local and covariant redefinition of the
symplectic current.
This suggests that such an extended symmetry
algebra cannot be realized as symmetries on the phase space of vacuum
general relativity at null infinity, and that the corresponding asymptotic
charges are ill-defined. However, a possible loophole in the argument
is the possibility that symplectic current may not need to be covariant
in order to have a covariant symplectic form.
We also show that the extended algebra does not have a preferred subalgebra of translations
and therefore does not admit a universal definition of Bondi 4-momentum.

\end{abstract}

\maketitle
\tableofcontents

\newpage

%%========================================================
\section{Introduction and Summary}
\label{sec:intro}

The asymptotic symmetry group at null infinity of asymptotically-flat
spacetimes in general relativity is normally considered to be the
infinite-dimensional Bondi-Metzner-Sachs (BMS) group. Associated with
the Lie algebra of the BMS group there are an infinite number of
charges and fluxes (due to gravitational radiation) at null infinity
\cite{BBM, Sachs:1962wk, Sachs:1962zza, Penrose, Geroch-asymp, GW,
  Ashtekar:1981bq, WZ}. Recently, these charges and fluxes have been
related to soft graviton theorems
\cite{He:2014laa,Strominger:2014pwa,Kapec:2015vwa,Avery:2015gxa,Campiglia:2015kxa,Campoleoni:2017mbt},
gravitational memory effects
\cite{He:2014laa,Strominger:2014pwa,Pasterski:2015tva,HIW,Mao:2017wvx,Pate:2017fgt,Chatterjee:2017zeb}
and potentially black hole information loss
\cite{Hawking:2016msc,Strominger:2017aeh,Hawking:2016sgy,Strominger:2017zoo}. 

There have been attempts to extend these relations to include the
subleading soft theorems, in particular, the one proposed by Cachazo
and Strominger in \cite{Cachazo:2014fwa}, by enlarging the
gravitational phase space to give additional asymptotic symmetries.
To understand the nature of these proposed enlargements consider,
briefly, the structure of the BMS algebra (see \cref{sec:asymp-flat}
for details). The BMS algebra contains an infinite-dimensional
extension of the usual translations --- known as
\emph{supertranslations} --- as well as (many equivalent subalgebras
of) infinitesimal Lorentz transformations. These Lorentz
transformations are conformal Killing fields on
\(2\)-sphere cross-sections of null infinity which are smooth
everywhere. It was proposed by Barnich and Troessaert
\cite{Barnich:2009se,Barnich:2010eb} (see also \cite{Kapec:2014opa})
that the BMS algebra should be extended to include the entire
infinite-dimensional Virasoro algebra, which consists of all
\emph{local} conformal Killing fields on a \(2\)-sphere\footnote{These
  symmetries are often called {\it superrotations}
  \cite{Barnich:2009se,Barnich:2010eb},
  although more recently that term has come to be used for the smooth infinitesimal
  diffeomorphisms on the 2-sphere \cite{Strominger:2017zoo}.
Another terminology for the smooth diffeomorphisms is {\it
  super-Lorentz} transformations, with the odd parity ones being
called superrotations and the even parity ones being called
superboosts \cite{Cnew}.}  . The vector
fields in the Virasoro algebra which are not Lorentz vector fields are
necessarily singular at isolated points on a \(2\)-sphere. An
alternative proposal by Campiglia and Laddha \cite{Campiglia:2014yka,
  CL} was to extend the Lorentz transformations by including all
smooth infinitesimal diffeomorphisms on a \(2\)-sphere. The
conservation law (i.e. Ward identity) for the charges at null infinity
corresponding to such extensions is then claimed to be equivalent to
the subleading soft theorem of \cite{Cachazo:2014fwa}.\\ 

In this paper, we use the symplectic formalism for general relativity to investigate whether such extensions of the asymptotic symmetries algebra have well-defined charges at null infinity. Since the Virasoro vector fields are in general singular, it is tricky to apply the usual symplectic formalism to such symmetries. Instead, we analyze the second extension of smooth 2-sphere diffeomorphisms \cite{Campiglia:2014yka, CL} mentioned above.

The main quantity of interest in our analysis is the \emph{symplectic current} derived from the Lagrangian of general relativity (\cref{symsec}). The symplectic current is a local and covariant \(3\)-form \(\df\omega(\tilde g; \delta_1 \tilde g, \delta_2\tilde g)\) which is an antisymmetric bilinear in two perturbations of the metric \(\delta_1\tilde g_{ab}\) and \(\delta_2\tilde g_{ab}\) (we have used a ``tilde'' to denote quantities in the physical spacetime as opposed to ones in the Penrose conformal completion). If suitable asymptotic conditions are satisfied then the symplectic current has a finite limit to null infinity. Then, the integral of the symplectic current over null infinity gives a symplectic form on the phase space of general relativity. If one of the perturbations, say \(\delta_2 \tilde g_{ab}\), is taken to be the perturbation generated by some asymptotic symmetry, then the symplectic form gives an expression for the generator of that symmetry on phase space. Note that the crucial aspect of the above is that the symplectic current must have a finite limit to null infinity, otherwise the generator would not be defined.

It is well-known that for asymptotically-flat spacetimes the
symmetries in the usual BMS algebra have well-defined generators in
the sense described above \cite{Ashtekar:1981bq,WZ}. We are interested
in whether generators corresponding to the extension of the BMS
algebra by all diffeomorphisms of a \(2\)-sphere exist. We show that
the answer is no: the symplectic current of general relativity
diverges in the limit to null infinity, in general, when one of the
perturbations is generated by an extended BMS symmetry (which is not a
BMS symmetry). This divergence was also previously encountered in the
computations of Compère, Fiorucci and Ruzziconi \cite{Cnew}.  
This divergence suggests that the generators of such extended
symmetries may not exist on the phase space at null infinity, and the
corresponding charges and fluxes may be also ill-defined.\\ 

A loophole in this argument is that one can exploit an ambiguity in
the symplectic current to render it finite in the limit to null infinity
(see Ref.\ \cite{Freidel:2019ohg} for a general discussion of such renormalization
in a different context).  The ambiguity is of the form $\df \omega \to \df \omega + 
d [\delta_{1}\df Y (\tilde{g};\delta_{2} \tilde{g}) - (1 \leftrightarrow 2)]$, for some two-form $\df Y$ which constructed out of the
dynamical fields and their variations \cite{WZ}.
Recently, Compère, Fiorucci and Ruzziconi
have shown that one can indeed obtain a finite symplectic current
using this method, and they find expressions for charges corresponding
to all the symmetries of the extended algebra, including the general
2-sphere diffeomorphisms \cite{Cnew}.

However, as noted by the authors themselves, their prescription
relies on a particular choice of coordinates
with the result that the two form $\df Y$ and the
final, finite symplectic current are not local, covariant function of
the dynamical fields.  Thus, it is not clear that the expressions
obtained in Ref.\ \cite{Cnew} for charges are unique. For instance, if
one repeated the construction using Newman-Unti coordinates instead of Bondi
coordinates, it is not clear if equivalent results would be obtained.
We will show that one cannot eliminate the divergences in the
symplectic current by exploiting the ambiguity in a local and
covariant manner.

This result suggests that the general 2-sphere diffeomorphisms do not
give rise to well defined charges and fluxes. However, a possible
loophole is that requiring that all the quantities in the construction
be local and covariant \cite{WZ} is too strong a restriction, and
instead
one should only impose this requirement on physically measurable quantities.
For example it
might be possible that the presymplectic form (obtained by integrating the
presymplectic current over a Cauchy surface) may be independent of the
arbitrary choice of coordinate system used in Ref.\ \cite{Cnew},
despite the fact that the presymplectic current $\df \omega$ does
depend on this choice.  It would be interesting to investigate this
possibility further, but we do not do so in this paper.

   The remainder of this paper is organized as follows. In
   \cref{sec:asymp-flat}, we review the definition of asymptotic
   flatness and show how the BMS algebra emerges as the asymptotic
   symmetry algebra of asymptotically-flat spacetimes. In \cref{sec:F-ext}, we consider the extended phase space proposed in \cite{CL} which leads to an extension of the BMS algebra to include arbitrary infinitesimal diffeomorphisms of a \(2\)-sphere. In \cref{symsec}, we show that the symplectic current evaluated on these extended symmetries diverges in the limit to null infinity. We also show
   that any local and covariant ambiguities in the symplectic current
   cannot get rid of this divergent behavior. We consider other issues associated with this extension of the BMS algebra in \cref{sec:issues}. We end by summarizing our main conclusions in Sec.~\ref{conc}. In \cref{sec:BS}, we
   construct suitable coordinate systems near \(\scri\) and show that
   the conformal factor and the unphysical metric at null infinity can be chosen universally in the class of asymptotically-flat
   spacetimes. In \cref{sec:trans-ideal}, we show that the extension of
   the BMS algebra by all diffeomorphisms of a \(2\)-sphere does not contain any preferred translation subalgebra.

%%..............................................................
\subsection{Notation and conventions}
\label{sec:not}

We follow the conventions of Wald \cite{Wald-book} throughout. Tensors
on spacetime will be denoted by Latin indices \(a,b,c,\ldots\). We
will frequently use an index-free notation for differential forms and
denote then by a bold-face, e.g. \(\df\omega \equiv \omega_{abc}\) is
the \(3\)-form symplectic current. Tensors on the physical spacetime
will be denoted by a ``tilde'' while those on the conformal completion
(unphysical spacetime) will not have a ``tilde'', e.g. \(\tilde
g_{ab}\) is the physical metric while \(g_{ab}\) is the unphysical
metric in the conformal completion.  Indices on unphysical, unbarred
quantities will be raised and lowered with the unphysical metric, for
example $n_a n^a = g^{ab} n_a n_b$.

%%============================================================  
\section{Asymptotic flatness at null infinity and the BMS algebra}
\label{sec:asymp-flat}

In this section, we review the definition of asymptotically-flat spacetimes and show how the BMS algebra arises as the asymptotic symmetry algebra at null infinity.\\

\subsection{Definition and properties of asymptotic flatness at null infinity}
\label{sec:dp}

A \emph{physical} spacetime $(\tilde{M},\tilde{g}_{ab})$, satisfying
the vacuum\footnote{For non-vacuum spacetimes, the definition of asymptotic flatness includes
  a fourth condition, that
  the physical stress-energy tensor ${\tilde T}_{ab}$ satisfies
  \be
     {\tilde T}_{ab} = \Omega^2 T_{ab}
     \label{eq:st}
     \ee
     for some tensor 
$T_{ab}$ which is smooth on $M$ including at $\scri$.  All of the results in this paper generalize to the non-vacuum case, except for the
discussion of the presymplectic current in Sec.\ \ref{symsec}, which is specialized to vacuum general relativity.}
Einstein equation \(\tilde G_{ab} = 0
\), is asymptotically-flat at null infinity if there exists another \emph{unphysical} spacetime $(M,g_{ab})$ with a boundary $\scri = \partial M$ and an embedding of \(\tilde M\) into \(M\),\footnote{We use the standard convention whereby the physical spacetime \(\tilde M\) is identified with its image in \(M\) under the embedding.} such that
\begin{enumerate}
\item there exists  and a smooth function $\Omega$ (the \emph{conformal factor}) on $M$ satisfying
    \begin{subequations}\begin{align} 
    \Omega = 0 \text{ on } \scri &\eqsp \nabla_{a} \Omega \text{ is nowhere vanishing on } \scri, \\
    g_{ab} = \Omega^2 \tilde g_{ab}& \text{ is smooth on } M \text{ including at } \scri.
    \end{align}\end{subequations}
\item $\scri$ is topologically $\bb R \times \bb S^2$.
\item Defining the vector field
  \be
  \label{eq:nadef}
  n_a = \nabla_a \Omega,
  \ee
 then the vector field \(\omega^{-1} n^a\) is complete on \(\scri\)
 for any smooth function \(\omega\) on \(M\) such that \(\omega > 0\)
 on \(M\) and \(\nabla_a(\omega^4 n^a) = 0\) on
 \(\scri\).\footnote{Since we will primarily be interested in the
   asymptotic symmetry algebra, and not the symmetry group, we will
   not need the completeness condition on \(\scri\).}
\end{enumerate}
For detailed expositions on the motivations for this definition, we
refer the reader to Refs.\ \cite{Geroch-asymp,Wald-book}. Note that the smoothness conditions on the unphysical spacetime can be significantly weakened --- we can allow \(M\) to have a \(C^3\)-differential structure and \(g_{ab}\) to be twice-differentiable at the boundary \(\scri\).

Using the conformal transformation relating the unphysical Ricci
tensor \(R_{ab}\) to the physical Ricci tensor \(\tilde R_{ab}\) (see
Appendix~D of \cite{Wald-book}), the vacuum Einstein equation can be written as
\be\label{eq:S-ee}
    S_{ab} = - 2 \Omega^{-1} \nabla_{(a} n_{b)} + \Omega^{-2} n^c n_c
    g_{ab}\,,
\ee
where \(S_{ab}\) is given by
\be\label{shoten}
    S_{ab} = R_{ab} - \tfrac{1}{6} R g_{ab}.
\ee
It follows immediately from \cref{eq:S-ee} and from
the assumed smoothness of
\(\Omega\) and of the unphysical metric \(g_{ab}\) at \(\scri\) that
\(n_a n^a \vert_\scri = 0\).  Hence \(\scri\) is a smooth null
hypersurface in \(M\) with normal \(n_a = \nabla_a \Omega\) and the
vector field \(n^a = g^{ab} n_b\) is a null geodesic generator of
\(\scri\).

Next, we write $n^a n_a = \chi \Omega$, where $\chi$ extends
smoothly to $\scri$, so that \cref{eq:S-ee} yields on $\scri$ that
\be
2 \nabla_{(a} n_{b)} = \chi g_{ab}.
\label{ghg}
\ee
Under a change of the conformal factor of the form
\be
\Omega \mapsto \omega \Omega, \ \ \ \ \ g_{ab} \mapsto \omega^2 g_{ab},
\label{eq:conformalfreedom}
\ee
where \(\omega\) is smooth in \(M\) and is nowhere vanishing on $\scri$, we have that $\chi$ transforms
on $\scri$ as $\chi \to ( \chi + 2 \lie_n \ln \omega)/\omega$.  Hence we can
choose $\omega$ to make $\chi=0$ \cite{Geroch-asymp,Wald-book},
which yields from \cref{ghg} the 
\emph{Bondi condition} 
\be\label{eq:Bondi-cond}
	 \nabla_{a} n_{b}|_{\scri}=\nabla_{a}\nabla_{b}\Omega|_{\scri}=0,
\ee
as well as
\be \label{eq:nn-cond}
    n_{a}n^{a} = O(\Omega^{2}).
\ee
The remaining freedom in the conformal factor is of the form
\cref{eq:conformalfreedom} with
\be\label{eq:conf-freedom}
    \omega\vert_\scri > 0 \eqsp \lie_n \omega\vert_\scri = 0.
\ee

Let \(q_{ab}\) the pullback of \(g_{ab}\) to \(\scri\). This defines a degenerate metric on \(\scri\) such that
\be\label{eq:q-cond}
    q_{ab} n^b = 0 \eqsp \lie_n q_{ab} = 0,
\ee
where the second condition follows from \cref{eq:Bondi-cond}. Thus, \(q_{ab}\) defines a Riemannian metric on the space of generators of \(\scri\) which is diffeomorphic to \(\bb S^2\).

A priori, the conformal completion depends on the physical spacetime \((\tilde M, \tilde g_{ab})\) under consideration. However, if \((M, g_{ab}, \Omega)\) and \((M', g'_{ab}, \Omega')\) are the unphysical spacetimes corresponding to \emph{any} two asymptotically-flat physical spacetimes, then \(M'\) can be identified with \(M\) using a diffeomorphism such that \(\scri'\) maps to \(\scri\), \(\Omega' = \Omega\) in a neighborhood of \(\scri\) and \(g'_{ab}\vert_\scri = g_{ab}\vert_\scri\) \cite{GW}. This can be shown by setting up a suitable geometrically-defined coordinate system in a neighborhood of \(\scri\) and identifying the two unphysical spacetimes in these coordinates; we defer the details to \cref{sec:BS}. Here we emphasize that the choice of coordinate system used is largely irrelevant. The only essential ingredients used in the identification are 
\begin{enumerate}
    \item \(\scri\) is a null smooth surface in the unphysical spacetime \(M\).
    \item The freedom in the choice of the conformal factor \(\Omega\)
      given by \cref{eq:conf-freedom}.
    \item The space of null generators of \(\scri\) is topologically
      \(\bb S^2\) and thus has a \emph{unique} conformal class of
      metrics up to diffeomorphisms.
\end{enumerate}
As discussed above, the first two ingredients
follow directly from the smoothness
requirements in the definition of asymptotic flatness and the Einstein
equation at \(\scri\). The third fact is a special case of the
\emph{uniformization theorem} (for instance see Ch.~8 \cite{Bieri})
and plays a crucial role on \(\scri\).\footnote{The uniformization
  theorem is a \emph{global} result depending on the topology of the
  \(2\)-dimensional space. Locally, all metrics of a particular
  signature on a \(2\)-surface are conformally-equivalent, Problem~2
  Ch.~3 \cite{Wald-book}.} As we will show below this fact leads
directly to the BMS algebra at \(\scri\) with the Lorentz algebra
being a subalgebra (instead of all diffeomorphisms of \(\bb
S^2\)). Further, it is also essential in the definition of a News
tensor characterizing the presence of radiation at \(\scri\) (see
Theorem~5 of \cite{Geroch-asymp}).\footnote{Even in spacetime dimensions \(d > 4\), to have well-defined Bondi mass and News tensor it appears essential to additionally assume that the metric \(q_{ab}\) on the \((d-2)\)-dimensional space of generators of \(\scri\) is conformal to a compact space of constant curvature \cite{Ish-Holl,Hol-Th,HIW}.}

As a result of this identification, we can work on a single manifold \(M\) with boundary
\(\scri\) and treat \(\Omega\) and \(g_{ab}\vert_\scri\) as
\emph{universal} within the entire class of asymptotically-flat
spacetimes, in the sense that they can be chosen to be independent of
the choice of the physical spacetime.
Specifically, fix a metric $g_{0\,ab}$ on $\scri$ and a conformal
factor $\Omega_0$ in a neighborhood ${\cal N}$ of $\scri$, and define
the field configuration space\footnote{Sometimes called the
pre-phase space or space of field histories \cite{Harlow:2019yfa}.}
\be
\Gamma_0 = \left\{ (M,g_{ab},\Omega) \right| \left. g_{ab \, |\scri} =
g_{0\,ab \, |\scri},  \ \ \ \Omega =
\Omega_0\  {\rm on}\  {\cal N}, \ \ \
	 \nabla_{a}\nabla_{b}\Omega|_{\scri}=0\right\}.
\label{eq:fcs}
  \ee
Note that we include the Bondi condition (\ref{eq:Bondi-cond}) in this definition.
Not all asymptotically flat unphysical metrics lie in $\Gamma_0$, but
for each one there corresponds an element of $\Gamma_0$ related to it
by a diffeomorphism and a conformal rescaling of the form (\ref{eq:conformalfreedom}),
as we show in Appendix \ref{sec:BS}.

\subsection{Review of derivation of the Bondi-Metzner-Sachs symmetry algebra}

The asymptotic symmetries at $\scri^+$ of the field configuration
space $\Gamma_0$ are the infinitesimal   
diffeomorphisms generated by vector fields \(\xi^a\) in \(M\) which
extend smoothly to \(\scri\), and whose pullbacks preserve the
asymptotic flatness conditions and map $\Gamma_0$ into itself, modded
out by the trivial 
diffeomorphisms whose asymptotic charges vanish \cite{WZ}.

Consider a one-parameter family of asymptotically-flat physical metrics \(\tilde g_{ab}(\lambda)\) where \(\tilde g_{ab} = \tilde g_{ab}(\lambda = 0)\) is some chosen background spacetime. Define the physical metric perturbation \(\tilde \gamma_{ab}\) around the background \(\tilde g_{ab}\) by
\be\label{eq:phys-pert}
    \tilde \gamma_{ab} = \delta \tilde g_{ab} = \lb. \frac{d}{d\lambda} \tilde g_{ab}(\lambda) \rb\vert_{\lambda = 0}
\ee
We will use ``\(\delta\)'' to denote perturbations of other quantities defined in a similar way.

Now let \(g_{ab}(\lambda)\) and \(\Omega(\lambda)\) be one-parameter family of unphysical metrics and conformal factors corresponding to the conformal completions of the physical metrics \(\tilde g_{ab} (\lambda)\). As discussed above, since the conformal factor \(\Omega\) is universal we have \(\delta \Omega = 0\) and \(\delta n_a = 0\). The unphysical metric perturbation is then
\be\label{ond}
	\delta g_{ab} = \gamma_{ab} = \Omega^2 \tilde \gamma_{ab}\,,
\ee
where \(\gamma_{ab}\) is smooth on \(M\) and extends smoothly to \(\scri\). Since the unphysical metric at \(\scri\) is universal, we have that
\be\label{need}
	\delta g_{ab}\vert_\scri = \gamma_{ab}\vert_\scri = 0 \implies \gamma_{ab} = \Omega \tau_{ab}\,,
\ee
for some tensor \(\tau_{ab}\) which extends smoothly to \(\scri\). Further, perturbing the Bondi condition (\ref{eq:Bondi-cond}) we can show that \cite{WZ}
\be\label{cond}
  \tau_{ab}n^{b} = \Omega \tau_{a}\,,
\ee
for some $\tau_{a}$ which extends smoothly to \(\scri\). Thus, the
unphysical metric perturbations \(\gamma_{ab}\) tangent to the field
configuration space $\Gamma_0$ of \cref{eq:fcs}
satisfy
\begin{subequations}
\label{eq:asymp-flat-conds}
  \begin{eqnarray}
\label{eq:asymp-flat-conds1}
    \gamma_{ab} &=& \Omega\tau_{ab},\\
\label{eq:asymp-flat-conds2}
    \gamma_{ab}n^b &=& \Omega^2 \tau_a.
  \end{eqnarray}
  \end{subequations}

Now consider the physical metric perturbation $\lie_\xi \tilde g_{ab}$ corresponding to an infinitesimal diffeomorphism generated by a vector field \(\xi^a\). The corresponding unphysical metric perturbation is
\be\label{eq:unphys-diffeo}
    \gamma^{(\xi)}_{ab} = \Omega^2 \lie_\xi \tilde g_{ab} =  \lie_\xi g_{ab} - 2 \Omega^{-1} n_c \xi^c g_{ab}.
    \ee
    For $\xi^a$ to be a representative of an infinitesimal asymptotic symmetry, the perturbation
    (\ref{eq:unphys-diffeo}) must satisfy the conditions 
(\ref{eq:asymp-flat-conds}).
First,  since \(\gamma^{(\xi)}_{ab}\) is smooth at
\(\scri\), we have from \cref{eq:unphys-diffeo} that \(n_a \xi^a \vert_\scri = 0\), that is, \(\xi^a\)
must be tangent to \(\scri\).  We define the function $\alpha_\xi$ by
\be
\label{eq:alphadef}
n_a \xi^a = \Omega
\alpha_{(\xi)},
\ee
where \(\alpha_{(\xi)}\) extends
smoothly to \(\scri\). This yields
\be\label{eq:xi-cond}
	\gamma^{(\xi)}_{ab} = \lie_\xi g_{ab} - 2 \alpha_{(\xi)} g_{ab},
\ee
and contracting with $n^a n^b$ and using Eqs.\ (\ref{eq:asymp-flat-conds}) and (\ref{eq:nn-cond}) gives that
\be
n^a n^b \nabla_a \xi_b = O(\Omega^2).
\label{eq:cfd}
\ee

Next, contracting \cref{eq:xi-cond} with \(n^b\) and using Eqs.\ (\ref{eq:Bondi-cond}) and (\ref{eq:alphadef}) gives
\be\label{eq:n-diffeo0}
    n^b\gamma^{(\xi)}_{ab} = n^b \nabla_b \xi_a - \xi^b \nabla_b n_a - \alpha_{(\xi)} n_a + \Omega \nabla_a \alpha_{(\xi)},
\ee
where we have used \(\nabla_a n_b = \nabla_b n_a\) from \cref{eq:nadef}. From Eqs.~(\ref{eq:asymp-flat-conds}) the
left-hand-side of \cref{eq:n-diffeo0} must vanish at \(\scri\), which gives
\be\label{eq:n-diffeo}
%    \lb. n^b\gamma^{(\xi)}_{ab} \rb\vert_\scri = 0 \implies 
    \lb. \lie_\xi n^a \rb\vert_\scri = - \lb. \alpha_{(\xi)} n^a \rb\vert_\scri.
\ee
Similarly the constraint (\ref{eq:asymp-flat-conds2}) gives using Eqs.\ (\ref{eq:nadef}),
(\ref{eq:nn-cond}), (\ref{eq:alphadef}), (\ref{eq:cfd}) and (\ref{eq:n-diffeo0})
\be\label{eq:nn-diffeo}
    n^a n^b \gamma^{(\xi)}_{ab} = O(\Omega^2) \implies \lb. \lie_n \alpha_{(\xi)} \rb\vert_\scri = 0.
\ee
Finally, the pullback of \cref{eq:xi-cond} to $\scri$ implies that
\be\label{needs}
    \lb. \gamma^{(\xi)}_{ab} \rb\vert_\scri = 0 \implies 
	\lb. \lie_\xi q_{ab} \rb\vert_\scri = 2 \alpha_{(\xi)} q_{ab}\,.
\ee
Thus, representatives of asymptotic symmetries for $\Gamma_0$ on \(\scri\) are 
vector fields \(\xi^a\) which are tangent to \(\scri\) and satisfy on
\(\scri\) 
\begin{subequations}\label{eq:bms-cond}\begin{align}
    \lie_\xi n^a &= - \alpha_{(\xi)} n^a, \label{eq:xi-n-cond} \\
    \lie_\xi q_{ab} &= 2 \alpha_{(\xi)} q_{ab}, \label{eq:xi-q-cond}
\end{align}\end{subequations}
where the function \(\alpha_{(\xi)}\) is smooth and
\be
\lie_n \alpha_{(\xi)} = 0
\label{eq:bms-cond1}
\ee
on \(\scri\).

Next we need to mod out by trivial infinitesimal diffeomorphisms for which all
boundary charges vanish.  For the case of vacuum general relativity,
the trivial vector fields $\xi^a$ are those which vanish on $\scri$
\cite{WZ},
and so it follows that the symmetry algebra consists of intrinsic vector fields
$\xi^a$ on $\scri$ which satisfy the conditions 
(\ref{eq:bms-cond}) and (\ref{eq:bms-cond1}) on $\scri$.  
These conditions
are the
familiar ones defining the BMS algebra \(\mf b\)
\cite{Geroch-asymp,Ashtekar:1981bq}.

Finally we review some of the properties of this algebra.  Consider
vector fields of the form \(\xi^a\vert_\scri = f n^a\) for which
\(\lie_n f \vert_\scri = 0\).
It follows from Eqs.\ (\ref{eq:nn-cond}) and (\ref{eq:alphadef}) that we have
\(\alpha_{(\xi)}\vert_\scri = 0\).
It
is easy to verify that vector fields of this form generate an
infinite-dimensional abelian subalgebra \(\mf s\) of \(\mf
b\). Further, this subalgebra is invariant in the sense that the Lie
bracket of any element of \(\mf b\) with any element of
\(\mf s\) is again in \(\mf s\), that is \(\mf s\) is a Lie ideal of
\(\mf b\). This is the subalgebra of \emph{supertranslations}. From \cref{eq:xi-q-cond}, the factor algebra \(\mf b / \mf s\) is 
isomorphic to the algebra of smooth conformal Killing fields of
\(q_{ab}\) on \(\bb S^2\). Since the conformal class of metrics on
\(\bb S^2\) is unique up to diffeomorphisms, the algebra of smooth conformal Killing fields
of \emph{any} \(q_{ab}\) is isomorphic to the algebra of smooth
conformal Killing fields of the unit-metric on \(\bb S^2\), that is
the \emph{Lorentz algebra} \(\mf{so}(1,3)\). Thus, the BMS algebra has
the semi-direct structure \(\mf b = \mf{so}(1,3) \ltimes \mf s\). The
BMS algebra also contains a unique \(4\)-dimensional Lie ideal
\(\mf t\) which can be interpreted as translations (see
\cite{Sachs:1962zza} or Theorem 6 of \cite{Geroch-asymp}, also
\cref{sec:trans-ideal} below). The presence of this 
preferred subalgebra \(\mf t\) implies that the Bondi \(4\)-momentum
at \(\scri\) is unambiguously defined.

%%------------------------------------------------------------
\section{An extended field configuration space and extended algebra}
\label{sec:F-ext}

In this section we show that by weakening the universal structure near
\(\scri\) --- equivalently, by extending the class of allowed
metrics --- one obtains a bigger asymptotic symmetry algebra at
null infinity which includes all the smooth diffeomorphisms of a
\(2\)-sphere. This is the algebra proposed by Campiglia and Laddha \cite{Campiglia:2014yka,
  CL}.

\subsection{Extended field configuration space}

It is clear from the preceding section that to obtain any extension of
the BMS algebra, one must enlarge the class of metrics under
consideration.  One option might be to suitably weaken the definition of
asymptotic flatness.  
An alternative approach, which we follow here, is to enlarge the definition (\ref{eq:fcs}) of the field
configuration space by relaxing the requirement that the unphysical
metric evaluated on $\scri$ be universal.  The motivation for this
enlargement is questionable, since the new metrics that are being
added are related to metrics already included in the
space $\Gamma_0$ by diffeomorphisms and
by the conformal transformations (\ref{eq:conf-freedom}).
This issue is discussed further in Sec.\ \ref{sec:ccps} below.
Nevertheless, we shall proceed and consider the extended class of metrics 
proposed by Campiglia and Laddha \cite{CL}.

In the definition of the extended field configuration space, we will
continue to require that the unphysical metric \(g_{ab}\) be smooth
at \(\scri\).
We will also continue to choose the conformal factor \(\Omega\) so
that the Bondi condition (\ref{eq:Bondi-cond}) holds.
Now, if we are given an unphysical spacetime $(M, g_{ab}, \Omega)$,
we can define tensors ${\hat n}^a$ and ${\bar \varepsilon}_{abc}$ intrinsic to $\scri$
by
\begin{subequations}
  \begin{eqnarray}
    {\hat n}^a &=& {g^{ab} \nabla_a \Omega} \vert_\scri,\\
    \label{eq:3vol}
    \varepsilon_{abcd}\vert_\scri &=& 4 \varepsilon_{[abc} n_{d]} \vert_\scri ,
  \end{eqnarray}
  \end{subequations}
and by defining \(\bar \varepsilon_{abc}\) to be the pullback of
\(\varepsilon_{abc}\) to \(\scri\).\footnote{Note that
  \(\varepsilon_{abc}\) is ambiguous up to \(\varepsilon_{abc} \mapsto
  \varepsilon_{abc} + \alpha_{[ab} n_{c]} \) but this does not affect
  the pullback \(\bar\varepsilon_{abc}\).}
It follows from the Bondi condition (\ref{eq:Bondi-cond}) that
\be
\lie_{\hat n} {\bar \varepsilon}_{abc} =0.
\ee

We now fix a choice of
tensors ${\hat n}_0^a$, ${\bar \varepsilon}_{0\,abc}$ obtained in this
way, fix a choice of conformal factor $\Omega_0$ on a neighborhood
${\cal N}$ of $\scri$, and define the extended field configuration space $\Gamma_{\rm
  ext}$ to be [compare \cref{eq:fcs}]
\be
\Gamma_{\rm ext} = \left\{ (M,g_{ab},\Omega) \right| \left. {\hat n}^a =
      {\hat n}^a_0, \ \ \ \ {\bar \varepsilon}_{abc} = {\bar \varepsilon}_{0\,abc},  \ \ \ \Omega =
  \Omega_0\  {\rm on}\  {\cal N}, \ \ \
	 \nabla_{a}\nabla_{b}\Omega|_{\scri}=0\right\}.
\label{eq:fcs1}
  \ee
This is the definition proposed by Campiglia and Laddha \cite{CL}, in
which the \(3\)-volume form \(\bar\varepsilon_{abc}\) and the normal
\({\hat n}^a\) at \(\scri\) are universal.  Note that $\Gamma_0$ is a
proper subset of $\Gamma_{\rm ext}$, if we choose the fields ${\hat
  n}^a_0$ and ${\bar \varepsilon}_{0\,abc}$ to be those associated
with $g_{0\,ab}\vert_\scri$.

\subsection{Extended algebra}

We now derive the form of the symmetry algebra for the field
configuration space (\ref{eq:fcs1}).
Since the fields ${\bar \varepsilon}_{abc}$ and ${\hat n}^a$ are
universal their perturbations must vanish, so
\be\label{conds} \begin{split}
\delta \bar\varepsilon_{abc}|_{\scri} = 0 \implies g^{ab}\gamma_{ab}|_{\scri} =0 \implies g^{ab}\gamma_{ab} = \Omega \sigma\,,\\
\quad
\delta n^{a}|_{\scri} = 0 \implies \gamma_{ab}n^{b} = \Omega \chi_{a}\,,
\end{split}\ee
for some fields $\sigma$ and \(\chi_a\) which extend smoothly to \(\scri\). Perturbing the Bondi condition (\ref{eq:Bondi-cond}) we have
\be
    \delta (\nabla_a n_b)\vert_\scri = 0 \implies \lb. n^c \nabla_c \gamma_{ab} \rb\vert_\scri = 2 n_{(a} \chi_{b)},
\ee
and taking the trace and using \cref{conds} gives
\be
    \lb. n^a \chi_a \rb\vert_\scri = 0.
\ee
Thus, the unphysical metric perturbations \(\gamma_{ab}\) in the extended class satisfy
\be\label{eq:ext-conds}
    g^{ab} \gamma_{ab} = \Omega \sigma \eqsp \gamma_{ab}n^b = \Omega \chi_a \eqsp \lb. n^a \chi_a \rb\vert_\scri = 0 \eqsp  \lb. n^c \nabla_c \gamma_{ab} \rb\vert_\scri = 2 n_{(a} \chi_{b)},
\ee
which are weaker than the conditions (\ref{eq:asymp-flat-conds}) on
perturbations in the conventional definition (\ref{eq:fcs}).

To find the asymptotic symmetries of this extended class of
spacetimes, let \(\gamma^{(\xi)}_{ab}\) be the unphysical
perturbation (\ref{eq:unphys-diffeo}) generated by a diffeomorphism
along \(\xi^a\), as before.
Imposing the requirements (\ref{eq:ext-conds}) we find that $\xi^a$
still satisfies the conditions 
(\ref{eq:nn-diffeo}) and (\ref{eq:n-diffeo}).
But since the unphysical metric is no longer universal at \(\scri\),
\(\gamma^{(\xi)}_{ab}\vert_\scri\) is no longer required to vanish and
so the condition (\ref{needs}) no longer holds.\footnote{The last condition in \cref{eq:ext-conds} does not impose additional restrictions on \(\xi^a\) at \(\scri\).} Using \(g^{ab}\gamma^{(\xi)}_{ab}\vert_\scri = 0 \) we have instead
\be
    \lie_\xi \bar\varepsilon_{abc} = 3 \alpha_{(\xi)} \bar\varepsilon_{abc},
\ee
where \(\bar\varepsilon_{abc}\) is the \(3\)-volume element
(\ref{eq:3vol}) on \(\scri\).

Modding out by the trivial diffeomorphisms as
before\footnote{\label{footnote:trivial}The trivial vector fields are those for which
    the integral of symplectic current evaluated on $\delta {\tilde g}_{ab}$ and $\lie_\xi
    {\tilde g}_{ab}$ vanishes \cite{WZ}.  However, this quantity cannot be evaluated since
    the symplectic current diverges on $\scri$, as we show in the next
    section.  So we simply assume here that the trivial vector fields
    $\xi^a$ are those which vanish on $\scri$, as for the standard
    definition of field configuration space.  The symmetry algebra
    thus could change given a finite renormalized symplectic current;
    see Sec.\ \ref{sec:ccps} below for further discussion of this point.}, we thus find
that the extended BMS algebra \(\mf b_{\rm ext}\) \cite{CL} of
$\Gamma_{\rm ext}$ is generated by vector fields \(\xi^a\) on $\scri$ which are
tangent to \(\scri\) and which satisfy
\begin{subequations}\label{eq:bms-ext-cond}\begin{align}
    \lie_\xi n^a &= - \alpha_{(\xi)} n^a, \label{eq:xi-n-ext-cond} \\
    \lie_\xi \bar\varepsilon_{abc} & = 3 \alpha_{(\xi)} \bar\varepsilon_{abc}, \label{eq:xi-vol-ext-cond}
\end{align}\end{subequations}
where the function \(\alpha_{(\xi)}\) is smooth and satisfies \cref{eq:bms-cond1}.
The structure of this extended BMS algebra can be analyzed as
before. There is an infinite-dimensional abelian Lie ideal \(\mf s\)
of supertranslations as before. However as a direct consequence of
dropping \cref{eq:xi-q-cond} in favor of \cref{eq:xi-vol-ext-cond},
the factor algebra \(\mf b_{\rm ext}/\mf s\) is now isomorphic to the
Lie algebra \(\mf{diff}(\bb S^2)\) of \emph{all} smooth
infinitesimal diffeomorphisms of \(\bb S^2\). Hence we have \(\mf b_{\rm ext} =
\mf{diff}(\bb S^2) \ltimes \mf s \).\\ 

Thus, by weakening the universal structure near \(\scri\) --- equivalently, extending the class of allowed perturbations --- one obtains a bigger asymptotic symmetry algebra at null infinity which includes all the smooth diffeomorphisms of a \(2\)-sphere.

%%==============================================================
\section{The symplectic current of general relativity at null infinity}
\label{symsec}

In this section we evaluate the symplectic current of general relativity for the extended class of perturbations detailed in \cref{sec:F-ext}. We will show that any choice of symplectic current, which is local and covariant, necessarily diverges in the limit to \(\scri\).\\

\subsection{The symplectic current for general perturbations}

We briefly review the symplectic formalism for general relativity, though this formalism can be used for any local and covariant Lagrangian theory. The Lagrangian \(4\)-form for vacuum general relativity is given by
\be \label{gr}
    \df L = \tfrac{1}{16 \pi} \tilde{\df\varepsilon}_4~ \tilde{R} 
\ee
where $\tilde{R}$ and $\tilde{\df\varepsilon}_4 \equiv
\tilde\varepsilon_{abcd}$ are the Ricci scalar and the volume
\(4\)-form, respectively, in the physical spacetime. The variation of
the Lagrangian is
\be\label{eq:var-L}
    \delta \df L = - \tfrac{1}{16\pi} \tilde G^{ab} \delta \tilde g_{ab} ~\tilde{\df\varepsilon}_4 + d \df\theta(\tilde g; \delta \tilde g)
\ee
where \(\tilde G_{ab}\) is the physical Einstein tensor and the \emph{symplectic potential} \(3\)-form \(\df \theta\) is given by \cite{WZ}
\be\label{eq:theta}\begin{aligned}
	\df\theta \equiv \theta_{abc} & = \tfrac{1}{16 \pi}\tilde\varepsilon_{dabc} \tilde v^d \quad\text{with} \\[1.5ex]
	\tilde v^a & = \tilde g^{ae} \tilde g^{fh} \lb[ \tilde \nabla_f \tilde \gamma_{eh} - \tilde \nabla_e \tilde\gamma_{fh} \rb]. \\
\end{aligned}\ee
Note that the symplectic potential (\ref{eq:theta}) is not uniquely determined by the Lagrangian. We will investigate the ambiguities in the symplectic formalism in \cref{sec:amb}.

The \emph{symplectic current} \(3\)-form \(\df \omega\) is defined by\footnote{In general the integral of the symplectic current over a \(3\)-dimensional surface will be degenerate and hence only defines a \emph{presymplectic} form on the space of fields. We shall not be concerned with such issues and so continue to call \(\df\omega\) a symplectic current.}
\be\label{sympcurrent}
 \df\omega (\tilde{g}; \delta_{1}\tilde{g},\delta_{2}\tilde{g}) = \delta_{1} \df\theta (\tilde{g}; \delta_{2} \tilde{g}) - \delta_{2} \df\theta(\tilde{g}; \delta_{1}\tilde{g})\,.
 \ee
 Using \cref{eq:theta}, the symplectic current $\df \omega$ for general relativity can be written as
$\omega_{abc} = {\tilde \varepsilon}_{dabc} {\tilde w}^d/(16 \pi)$ with \cite{WZ}
\be\begin{aligned}\label{presym}
	\tilde w^a & = \tilde P^{abcdef} \tilde\gamma_{2bc} \tilde\nabla_d \tilde\gamma_{1ef} - [1 \lra 2], \\
    \tilde P^{abcdef} & = \tilde{g}^{ae}\tilde{g}^{fb}\tilde{g}^{cd} - \tfrac{1}{2}\tilde{g}^{ad}\tilde{g}^{be}\tilde{g}^{fc} - \tfrac{1}{2}\tilde{g}^{ab}\tilde{g}^{cd}\tilde{g}^{ef} - \tfrac{1}{2}\tilde{g}^{bc}\tilde{g}^{ae}\tilde{g}^{fd} + \tfrac{1}{2}\tilde{g}^{bc}\tilde{g}^{ad}\tilde{g}^{ef},
\end{aligned}\ee
where ``\([1 \lra 2]\)'' denotes the preceding expression with the
labels \(1\) and \(2\), labeling the perturbations, interchanged.\\

If the metric \(\tilde g_{ab}\), the perturbation \(\delta\tilde
g_{ab}\) and the vector field \(\xi^a\) generating an infinitesimal
diffeomorphism satisfy some suitable asymptotic conditions at
\(\scri\) so that the symplectic current \(\df\omega(\tilde g;
\delta\tilde g, \lie_\xi \tilde g)\) has a finite limit to \(\scri\),
then the integral of this symplectic current over \(\scri\) can be
used to define a generator of the symmetry corresponding to \(\xi^a\)
at null infinity \cite{WZ}. Note that we require the full \(3\)-form \(\df\omega\) to have a limit to \(\scri\) and not just its pullback or integral over some surfaces that limit to \(\scri\); the latter procedure would, in general, depend on the choice of surfaces used and would not be covariant.

We now investigate whether the symplectic current has a limit to
\(\scri\) when \(\xi^a\) is a generator of the extended BMS
algebra. To analyze the behavior of the symplectic current at $\scri$,
it is convenient to express the symplectic current in terms of
unphysical quantities which extend smoothly to $\scri$ and are
non-vanishing there in general.  Thus we define
\be\label{def}
	{\varepsilon}_{abcd} = \Omega^{4} \tilde{\varepsilon}_{abcd} \,, \quad  P^{abcdef} = \Omega^{-6} \tilde{P}^{abcdef} \,, \quad  \gamma_{ab} = \Omega^{2} \tilde{\gamma}_{ab}\,.
\ee
The relation between the physical derivative operator \(\tilde \nabla\) and unphysical derivative operator \(\nabla\) acting on any covector \(v_a\) is given by
\be\label{eq:der-conv}\begin{aligned}
    \tilde \nabla_a v_b &= \nabla_a v_b + C^c{}_{ab} v_c \\
    \text{with}\quad  C^c{}_{ab} &= 2 \Omega^{-1} \delta^c_{(a} n_{b)} - \Omega^{-1} n^c g_{ab}
\end{aligned}\ee
Inserting \cref{def} into \cref{presym}, using \cref{eq:der-conv}, one obtains:
\be\begin{aligned}\label{imp}
	 \df\omega \equiv \omega_{abc} & = \tfrac{1}{16 \pi}\varepsilon_{dabc} w^d \,, \quad \\[1.5ex]
	\text{with}\quad 
    w^a & = \Omega^{-2} P^{abcdef} \gamma_2{}_{bc} \nabla_d \gamma_1{}_{ef} + \Omega^{-3} \gamma_1^{ab} n_b \gamma_{2 c}{}^c - [1 \lra 2] \,.
\end{aligned}\ee\\

Note that if both perturbations \(\gamma_1{}_{ab}\) and
\(\gamma_2{}_{ab}\) are tangent to the standard field configuration space
(\ref{eq:fcs}) and
so satisfy the usual conditions (\ref{eq:asymp-flat-conds}), we have
\be
	w^a = P^{abcdef} \tau_2{}_{bc} \nabla_d \tau_1{}_{ef} +
        \tau_1{}^{ab} \tau_2{}_b + \tau_1{}^a \tau_{2 c}{}^c - [1 \lra
          2] + O(\Omega) \,,
\ee
and in this case the symplectic current is finite at $\scri$. In particular, when one of the perturbations is generated by an infinitesimal diffeomorphism corresponding to the BMS algebra, one can define the corresponding charges and fluxes following the procedure in \cite{WZ}.

\subsection{Divergence of the symplectic current on the extended phase space}

Now consider the case where one of the perturbations, say
$\gamma_{1ab}$, satisfies the usual set (\ref{eq:asymp-flat-conds}) of
conditions, while the other one, $\gamma_{2ab}$, lies in $\Gamma_{\rm
  ext}$ and so satisfies only the weaker set of conditions (\ref{eq:ext-conds}). If
in this case the symplectic current has a finite limit to \(\scri\)
then one can hope to define charges and fluxes for the extended BMS
algebra taking \(\gamma_{2ab} = \gamma^{(\xi)}_{ab}\) where
\(\xi^a\vert_\scri\) is an element in \(\mf b_{\rm ext}\). However, as
we now show, the symplectic current (\ref{imp}) in this case
necessarily diverges in the limit to \(\scri\), unless
\(\gamma_2{}_{ab}\)  also lies in the standard field configuration 
space $\Gamma_0$ [i.e.\ satisfies the
standard conditions (\ref{eq:asymp-flat-conds})], or the
perturbation \(\gamma_1{}_{ab}\) contains no
gravitational radiation, i.e., has vanishing perturbed News tensor. 

If the symplectic current \(\df\omega\) has a limit to \(\scri\) then its pullback to \(\scri\) will be proportional to \(\bar{\df\varepsilon}_3 ~ n_a w^a\) where \(\bar{\df\varepsilon}_3 \equiv \bar\varepsilon_{abc}\) is the \(3\)-volume element on \(\scri\). It will suffice for our purposes to show that \(n_a w^a\) does not have a limit to \(\scri\). Using Eqs.\ (\ref{cond}), (\ref{need}) for $\gamma_{1ab}$ and Eqs.\ (\ref{conds}) for $\gamma_{2ab}$ we have
\be\label{eq:n-omega}\begin{split}
    n_{a}w^{a} & = \Omega^{-1} \lb[ n_{a} P^{abcdef}\gamma_{2bc}\nabla_{d}\tau_{1ef} - n_{a} P^{abcdef}\tau_{1 bc}\nabla_{d}\gamma_{2ef}  \rb] \\
    &\quad + \chi_{2 a} \tau_{1}{}^{a} + \tfrac{1}{2}\sigma_2 \tau_{1}{}^{a} n_{a} -\tfrac{3}{2} \Omega^{-1} n^{a}\chi_{2a} \tau_{1 b}{}^b -\tfrac{1}{2} \Omega^{-2} n_{a} n^{a} \gamma_{2}{}^{bc} \tau_{1 bc} + \tfrac{1}{2} \Omega^{-1} n_{a} n^{a} \sigma_2 \tau_{1 b}{}^b
\,.
\end{split}\ee
Due to Eqs.\ (\ref{eq:nn-cond}) and (\ref{eq:ext-conds}) the second line in \cref{eq:n-omega} has a finite limit to \(\scri\). Now if \(n_a w^a\) has a finite limit to \(\scri\) then we must have \(\lim_{\to \scri} \Omega n_a w^a = 0\). From \cref{eq:n-omega} we have
\be\label{simp}
    \Omega n_a w^a =  n_{a}P^{abcdef}\gamma_{2bc}\nabla_{d}\tau_{1ef} - n_{a} P^{abcdef}\tau_{1bc}\nabla_{d}\gamma_{2ef} + O(\Omega),
 \ee
where \(O(\Omega)\) indicates terms which vanish in the limit to \(\scri\). Evaluating the first term of Eq.\ (\ref{simp}) at \(\scri\), using Eq.\ (\ref{eq:ext-conds}), we have
\be\begin{aligned}
    \lb. n_{a}P^{abcdef}\gamma_{2bc}\nabla_{d}\tau_{1ef} \rb\vert_\scri
    & = n^a \gamma_{2}{}^{bc} \nabla_c \tau_{1ab} - \tfrac{1}{2} n^a \gamma_{2}{}^{bc} \nabla_a \tau_{1 bc} \\
    & = - \tfrac{1}{2} \gamma_{2}{}^{bc} n^a \nabla_a \tau_{1 bc}.
\end{aligned}\ee
Similarly for the second term in Eq.\ (\ref{simp}) we have (from Eqs.\ (\ref{eq:asymp-flat-conds}), (\ref{eq:ext-conds}))
\be
    \lb. n_{a} P^{abcdef}\tau_{1bc}\nabla_{d}\gamma_{2ef} \rb\vert_\scri
    = - \tfrac{1}{2} \tau_1{}^{bc} n^a \nabla_a\gamma_{2 bc} = 0.
\ee
Thus Eq.\ (\ref{simp}) can be written as
\be\label{simp-more}
    \Omega n_a w^a =  - \tfrac{1}{2} \gamma_{2}{}^{ab} n^c \nabla_c \tau_{1 ab} + O(\Omega).
\ee

We now simplify the above expression further using the vacuum Einstein equation. Perturbing Eq.\ (\ref{eq:S-ee}), for the perturbation \(\gamma_1{}_{ab}\) we have (see also Eq.~(67) of \cite{WZ})
\be\label{eq:gamma1-eqn}
    \delta_1 S_{ab} \vert_\scri = 2 n_a \tau_{1 b} + 2 n_b \tau_{1 a} - n^c \nabla_c \tau_{1 ab} - n^c \tau_{1 c} g_{ab}.
\ee
Using Eq.\ (\ref{eq:gamma1-eqn}) in Eq.\ (\ref{simp-more}) to eliminate the derivative of \(\tau_{1ab}\), and Eqs.\ (\ref{eq:asymp-flat-conds}), (\ref{eq:ext-conds}) we get
\be
    \Omega n_a w^a = \tfrac{1}{2}\gamma_{2}{}^{ab}\delta_1 S_{ab} + O(\Omega).
\ee
Further, since $\gamma_{2}{}^{ab}$ is tangent to $\scri$ (from Eq.\ (\ref{conds})) we can replace \(S_{ab}\) by its pullback to \(\scri\) \(\bar S_{ab}\) to get
\be \label{oss}
    \Omega n_a w^a = \tfrac{1}{2}\gamma_{2}{}^{ab}\delta_1 \bar S_{ab} + O(\Omega)
\ee

Now for a asymptotically flat spacetime the News tensor on \(\scri\) is defined by
\be
    N_{ab} = \bar S_{ab} - \rho_{ab},
\ee
where \(\rho_{ab}\) is the unique symmetric tensor field on \(\scri\)
constructed from the (usual) universal structure at \(\scri\) in
Theorem~5 of \cite{Geroch-asymp}. Thus, for the perturbation
$\gamma_{1\,ab}$ we have \(\delta_1 \rho_{ab} = 0\) and we can replace \(\delta_1 \bar S_{ab}\) in \cref{oss} with \(\delta_1 N_{ab}\) to get
\be\label{eq:n-omega-News}
    \Omega n_a w^a = \tfrac{1}{2}\gamma_{2}{}^{ab}\delta_1 N_{ab} + O(\Omega)
\ee
For general perturbations \(\gamma_{1 ab}\) the perturbed News \(\delta_1 N_{ab}\) does not vanish on \(\scri\), indicating the presence of (linearized) gravitational radiation,
although it is subject to the constraints
\be
g^{ab} \delta_1 N_{ab} = 0, \ \ \ n^a \delta_1 N_{ab}=0.
\ee
If the quantity (\ref{eq:n-omega-News}) vanishes for all perturbations $\gamma_{1\, ab}$, then $\gamma_{2\, ab}$ must be of the form
$\alpha g_{ab} + n_{(a} v_{b)} + O(\Omega)$, but it then follows from Eqs.\ (\ref{eq:ext-conds}) that $\gamma_{2\, ab} = O(\Omega)$.

We therefore conclude that 
\begin{enumerate}
    \item The symplectic current has a finite limit to \(\scri\) for
      \emph{all} perturbations \(\gamma_1{}_{ab}\) that are tangent to the
      standard phase space $\Gamma_0$, if and only if \(\gamma_{2
        ab}\vert_\scri = 0 \), that is, \(\gamma_{2 ab}\) also is tangent to
      $\Gamma_0$. In particular when \(\gamma_{2 ab} =
      \gamma^{(\xi)}_{ab}\) is a perturbation generated by an
      infinitesimal diffeomorphism \(\xi^a\), then
      \(\gamma^{(\xi)}_{ab}\vert_\scri = 0\) and thus
      \(\xi^a\vert_\scri\) is an element of the usual BMS algebra
      \(\mf b\) (see \cref{needs}).
    \item The symplectic current has a finite limit to \(\scri\) for any \(\gamma_2{}_{ab} = \gamma^{(\xi)}_{ab}\) generated by an infinitesimal diffeomorphism \(\xi^a\) in \(\mf b_{\rm ext}\) which is not in \(\mf b\), if and only if \(\gamma_1{}_{ab}\) has vanishing perturbed News, that is, \(\gamma_1{}_{ab}\) is non-radiating at \(\scri\).
\end{enumerate}

We emphasize that we have shown that (except in the cases discussed above) the limit to \(\scri\) of the symplectic current \(\df\omega\) as a \(3\)-form does not exist. That is, the symplectic current diverges as we approach \emph{any} point of \(\scri\) along \emph{any} curve in the unphysical spacetime independently of any choice of coordinates.\\

We now compare our result with the procedure used by Campiglia and
Laddha \cite{CL}, who obtained finite charges associated with
generators of the extended algebra.  Their procedure
can be described as follows. In the physical
spacetime pick some Bondi coordinate system \((r,u,x^A)\) near
\(\scri\). Consider the surfaces \(\Sigma_t\) given by \(t= u + r =
\text{constant}\) and integrate the symplectic current \(\df\omega\)
on \(\Sigma_t\) with the perturbation \(\delta_2\tilde g_{ab} =
\lie_\xi \tilde g_{ab}\) generated by some diffeomorphism in \(\mf
b_{\rm ext}\) and \(\delta_1 \tilde g_{ab}\) lying in $\Gamma_0$. This
integral can be rewritten as an integral over a \(2\)-sphere of \(u =
\text{constant}\) on \(\Sigma_t\). Then as \(u \to -\infty\) this
integral diverges linearly in \(u\) if the vector field \(\xi^a\) is
an element of \(\mf b_{\rm ext}\) which is not in the usual BMS
algebra \(\mf b\). To get a finite symplectic form for all symmetries
in \(\mf b_{\rm ext}\), Ref.\ \cite{CL} then imposes the boundary condition \(C_{AB} \sim 1/u^{1+\epsilon}\) along \emph{every} \(\Sigma_t\) where \(C_{AB}\) is a subleading piece of the physical metric on the \(2\)-spheres in Bondi coordinates. The symplectic form on \(\scri\) is then defined as the \(t\to\infty\) limit of this symplectic form on the surfaces \(\Sigma_t\). 

We note that, in contrast to our approach, the procedure used by \cite{CL} is not covariant. In particular their boundary condition \(C_{AB} \sim 1/u^{1+\epsilon}\) along \emph{every} \(\Sigma_t\) is not invariant under supertranslations (this was noted also by \cite{CL}). Thus, if this condition holds in one choice of Bondi coordinate system, it fails to hold in another Bondi coordinate system related to the first by a supertranslation. Similarly, this condition fails to hold if one instead integrates the symplectic current on some different family of surfaces which are supertranslated relative to their choice of \(\Sigma_t\). We also note that for the ``soft charge'' on \(\scri\) defined in \cite{Campiglia:2014yka,CL} to be finite one needs to impose \(C_{AB} \sim 1/|u|^{1+\epsilon}\) along \(\scri\) as \(u \to \pm\infty\) (the ``soft charge'' vanishes for elements of the Lorentz algebra \(\mf{so}(1,3)\) upon integration over the $2$-spheres and this restriction is not required). As is well-known \cite{HIW}, this implies that the memory effect in such spacetimes must vanish which is a severe restriction on the class of spacetimes.

%%--------------------------------------------------------------
\subsection{Ambiguities in the symplectic current}
\label{sec:amb}

As shown in the previous section, the symplectic current of a
perturbation in the standard phase space $\Gamma_0$
with any element of the extended BMS algebra (which is not in the usual BMS algebra) is not finite at \(\scri\). However, the symplectic current of general relativity is not uniquely determined by its Lagrangian, and it was claimed in \cite{Cnew} that the symplectic current can be made finite at \(\scri\) by a suitable choice of such an ambiguity. Since the computations of \cite{Cnew} are tied to a Bondi coordinate system, it is not apparent if their choice of the ambiguity is local and covariant. In this section we show that any ambiguity in the symplectic current, which is local and covariant cannot be used to make the symplectic current finite at \(\scri\), in general.\\

The symplectic potential \(\df\theta\) defined by \cref{eq:var-L} is ambiguous up to the addition of a local and covariant \(3\)-form \(\df X(\tilde g;\delta \tilde g)\) which is linear in \(\delta\tilde g\) and closed, i.e. \(d\df X = 0\). It can be shown quite generally \cite{W-closed} that such a closed form must be exact i.e. \(\df X(\tilde g; \delta\tilde g) = d \df Y(\tilde g ;\delta\tilde g)\) for some \(2\)-form $\df Y$ which is local and covariant and linear in \(\delta \tilde g\). Thus, the ambiguity in the symplectic potential is
\begin{equation}\label{ambg}
 \df\theta(\tilde g; \delta \tilde g) \mapsto \df\theta(\tilde g; \delta \tilde g) + d\df Y (\tilde g; \delta \tilde g)
 \end{equation} 
From \cref{sympcurrent}, the corresponding ambiguity in the symplectic current is given by\footnote{Note that the equations of motion are unaffected by the change \(\df L \mapsto \df L + d\df K\) in the Lagrangian. This does not affect the symplectic current since \(\delta_1 \delta_2 \df K - \delta_2 \delta_1 \df K = 0\).} 
\begin{equation}
\df\omega (\tilde g; \delta_1 \tilde{g}, \delta_2 \tilde{g}) \mapsto \df\omega(\tilde g; \delta_1 \tilde{g},\delta_2 \tilde{g}) + d\left[ \df Z(\tilde g; \delta_1 \tilde{g}, \delta_2 \tilde{g}) - \df Z(\tilde g; \delta_2 \tilde{g}, \delta_1 \tilde{g}) \right]
\end{equation}
where, for later convenience, we have defined the \(2\)-form
\be \label{eq:Z-defn}
\df Z(\tilde g; \delta_1 \tilde{g}, \delta_2 \tilde{g}) = \delta_1 \df Y(\tilde g; \delta_2 \tilde{g}) \,,
\ee
Note that it follows from \cref{eq:Z-defn} that any such \(\df Z\) must satisfy the condition
\be\label{cod}
 \delta_3 \df Z (\tilde g; \delta_1 \tilde{g}, \delta_2 \tilde{g}) - \delta_1 \df Z(\tilde g; \delta_3 \tilde{g}, \delta_2 \tilde{g}) = 0
 \ee
for \emph{arbitrary} perturbations $\delta_3 \tilde g$ of the metric,
even those that do not lie in $\Gamma_0$.

We first define the notion of a scaling dimension for tensors following \cite{Ish-Holl}. A tensor $L^{a\ldots}{}_{b\ldots}$ with $u$ upper and $l$ lower indices constructed out of the unphysical metric \(g_{ab}\) and the conformal factor \(\Omega\) is said to have a scaling dimension $s$, if under a scaling of the conformal factor $\Omega \mapsto \lambda \Omega$ and the metric \(g_{ab} \mapsto \lambda^2 g_{ab}\) by a constant \(\lambda\), we have $L^{a\ldots}{}_{b\ldots} \mapsto \lambda^{s-u+l} L^{a\ldots}{}_{b\ldots}.$ Note that the scaling dimension is independent of the tensor index positions and is additive under tensor products. One sees from this that the relevant scaling dimensions are: 
\be \label{scaling}
    \Omega: 1 \eqsp g_{ab}: 0 \eqsp \varepsilon_{abcd} : 0 \eqsp \nabla:-1 \eqsp n_{a}: 0
\ee

Since we are interested in the behavior of \(\df Z\) near $\scri$, it
is useful to write everything in terms of unphysical quantities which
are smooth at \(\scri\). Since $\df Z (\delta_1 \tilde{g} , \delta_2
\tilde{g})$ is linear in both physical metric perturbations, in terms
of the unphysical perturbations \(\gamma_{1 ab}, \gamma_{2 ab}\) we
must have
\be \label{ans}
    \df Z (\gamma_{1}, \gamma_{2}) = \sum_{p,q} \df W^{abcd e_1 \ldots e_p f_1 \ldots f_q} (\nabla_{e_1} \cdots \nabla_{e_p} \gamma_{1ab}) (\nabla_{f_1} \cdots \nabla_{f_q} \gamma_{2cd}) 
\ee
where $p$ and \(q\) (each ranging from \(0\) to some finite value)
count the number of derivatives of \(\gamma_{1ab}\) and
\(\gamma_{2ab}\), respectively. Here \(\df W^{abcd e_1 \ldots e_p f_1 \ldots f_q}\) are some local and covariant tensor-valued \(2\)-forms which are local functionals of the unphysical metric, the unphysical Riemann tensor and its derivatives and the conformal factor. The scaling dimension of $\gamma_{1ab} $ and $\gamma_{2ab}$ is $0$, and since the scaling dimension of the symplectic potential \(\df\theta\) is $-3$ it follows from \cref{ambg} that the scaling dimension of $\df Z$ is $-2$. Therefore, the scaling dimension of $\df W^{abcd e_1 \ldots e_p f_1 \ldots f_q}$ is $-2 + p +q$.

Now we analyze the possible forms of $\df W^{abcd e_1 \ldots e_p f_1
  \ldots f_q}$ that can appear in \cref{ans}. Note that our goal is to
find a \(\df Z\) that can get rid of the divergence in the symplectic
current in the limit to \(\scri\). From \cref{eq:n-omega-News} we see
that this diverging term depends analytically on the background
unphysical metric. Thus, in any candidate expression for \(\df Z\) of
the form (\ref{ans}), we can assume that \(\df W^{abcd e_1 \ldots e_p
  f_1 \ldots f_q}\) is an analytic functional of its
arguments.\footnote{We emphasize that \(\df W^{abcd e_1 \ldots e_p f_1
    \ldots f_q}\) being analytic in its functional dependence is
  unrelated to the analyticity of the unphysical metric on the
  spacetime manifold. We do not impose any analyticity conditions on
  the spacetimes under consideration.} Using the Einstein equation (\ref{eq:S-ee}), we can eliminate the covariant derivatives of \(n_a\) in favor of \(S_{ab}\) and its derivatives. Similarly, the unphysical Riemann tensor and its derivatives can be rewritten in terms of $S_{ab}$ and the Weyl tensor $C_{abcd}$ and their derivatives using
\be
    R_{abcd} = C_{abcd} + g_{a[c} S_{d]b} - g_{b[c} S_{d]a}
\ee
Thus any typical term in \(\df W^{abcd e_1 \ldots e_p f_1 \ldots
  f_q}\) can be schematically written in the form\footnote{Note that,
  by the peeling theorem (Theorem~11 \cite{Geroch-asymp}), for an
  asymptotically-flat spacetime, \(C_{abcd}\) vanishes and
  \(\Omega^{-1} C_{abcd}\) has a finite limit at \(\scri\). Thus, in
  \cref{ansatz} we can use \(\Omega^{-1} C_{abcd}\) instead; this only
  changes the last term in \cref{sc-dim} to \(-3 u\) and does not
  affect the rest of the argument. We use the Weyl tensor \(C_{abcd}\)
  since we allow the background spacetime to satisfy some ``extended''
  notion of asymptotic flatness for which the peeling theorem might not hold.}
\be \label{ansatz}
    \Omega^{v} \prod^{r}_{i=1} (\nabla)^{s_{i}} S_{ab} \prod_{j=1}^{u} (\nabla)^{t_{j}} (C_{abcd})\times (\text{terms with 0 scaling dimension})\,,
 \ee
where we have suppressed contractions with the metric \(g_{ab}\) for simplicity of notation. In the above \(r,u \geq 0 \) count the number of factors involving \(S_{ab}\) and \(C_{abcd}\) respectively and $s_{i}, t_{j} \geq 0$ count the number of derivatives occurring in each such term. Note that \(v\) is allowed to be negative. Comparing the scaling dimensions of \cref{ansatz} and $\df W^{abcd e_1 \ldots e_p f_1 \ldots f_q}$ gives
\be \label{sc-dim}
    -2 + p + q = v - \sum_{i}^{r} s_{i} - 2 r - \sum_{j}^{u} t_{j} - 2u \,.
\ee
From the above we see that \(v \geq -2\). Lets consider the ``most singular'' term where \(v = -2\), and thus $p = q = r = u=0$; this term does not contain any \(S_{ab}\) or \(C_{abcd}\) and has no derivatives of the perturbations \(\gamma_{1ab}, \gamma_{2ab}\). Then, \cref{ans} simplifies to the form 
\be \label{eq:Z-simp}
    \df Z (\gamma_{1}, \gamma_{2}) = \df W^{abcd} \gamma_{1ab} \gamma_{2cd} + O(\Omega^{-1})\,,
\ee
where $\df W^{abcd} = \Omega^{-2} \times (\text{terms with 0 scaling
  dimension})$. Recall that when \(\gamma_1{}_{ab}\) is a perturbation
in $\Gamma_0$ we have \(\gamma_1{}_{ab} = \Omega \tau_1{}_{ab}\) where \(\tau_1{}_{ab}\) is smooth and non-vanishing, in general, at \(\scri\). In this case, the ``most singular'' term we have considered in \cref{eq:Z-simp} diverges as \(\Omega^{-1}\) near \(\scri\). This is precisely the term one would need to cancel the diverging part of the symplectic current in \cref{eq:n-omega-News}.

 Let us now figure out what the \(2\)-form $\df W^{abcd}$ can be. Notice that $\Omega$ can only appear with a power $-2$ in the expression for $\df W^{abcd}$, in particular, any terms with $0$ scaling dimension that we need cannot be constructed by multiplying some powers of $\Omega$ with something with a negative scaling dimension. Since \(\df W^{abcd}\) must be local and covariant the only quantities available are $g_{ab}, \varepsilon_{abcd}$ and $n_{a}$ --- note that any derivatives of these will have negative scaling dimension. Using \cref{eq:nn-cond,eq:ext-conds} leads to just two possible terms which appear at order \(\Omega^{-2}\), in terms of which we can write \cref{eq:Z-simp} as
\be\label{eq:Z-final}
    \df Z (\gamma_{1}, \gamma_{2}) \equiv Z_{ab} (\gamma_{1}, \gamma_{2}) = \Omega^{-2} (A \varepsilon_{ab}{}^{cd} + B \delta_{[a}^c \delta_{b]}^d) g^{ef} \gamma_{1ce} \gamma_{2df} + O(\Omega^{-1})
\ee
where \(A,B\) are some constants. Since we have only computed the \(\df Z\) up to terms of \(O(\Omega^{-1})\), our consistency condition \cref{cod}, must also hold to this order. However, it is easy to verify that \cref{eq:Z-final} fails to satisfy this condition since $\delta_{3} g_{ab}|_{\scri} \neq 0$ for an arbitrary perturbation in the extended class of perturbations. That is, there does not exist an ambiguity \(\df Y\) in the symplectic potential such that \cref{eq:Z-final} is of the form \cref{eq:Z-defn}. Thus we conclude that \emph{any} choice of the symplectic current for general relativity, which is local and covariant, must diverge in the limit to \(\scri\), in general, when at least one of the perturbations is taken to be in the extended class of allowed perturbations.

\section{Other Issues}
\label{sec:issues}

Since our covariance arguments are not airtight, as discussed in the
Introduction, we now consider some other arguments for
and against the extension of the BMS algebra.  We focus on two
specific issues: the desirability of having a definition of Bondi
4-momentum and the freedom in choosing a field configuration space.

\subsection{Existence of Bondi four-momentum}
\label{sec:bondi4}

The standard BMS algebra \(\mf b \) contains a preferred four dimensional subalgebra of translations, associated
with the existence of Bondi 4-momentum.  By contrast, the extended
algebra \(\mf b_{\rm ext}\) does not, as we show explicitly in
Appendix \ref{sec:trans-ideal}.  Therefore, there is no natural
universal definition of Bondi 4-momentum in any context where \(\mf
b_{\rm ext}\) is the asymptotic symmetry algebra.  This lack of a
definition of Bondi 4-momentum would seem to be a difficulty for any
physical interpretation of the extended algebra.

However, the notion of Bondi 4-momentum would still apply in
the context of the symmetry algebra \(\mf b_{\rm ext}\), but
in a solution-dependent manner.  Specifically, given a solution
$(M,g_{ab},\Omega)$, one can define the field configuration space
(\ref{eq:fcs}) associated with that solution, and from it obtain an
associated translation subgroup of \(\mf b_{\rm ext}\) and corresponding
4-momentum charge.  The 4-momenta associated with two different
solutions need not be comparable, as in general they would lie in
different spaces.  This status of 4-momentum in the extended algebra
would be analogous to the status of angular momentum 
in the standard BMS context.  There, stationary solutions determine preferred
Poincar\'e subalgebras of the BMS algebra, with associated linear and
angular momentum charges, but the angular momentum charges associated
with two different stationary solutions need not be comparable as they
live in different spaces.

\subsection{Choice of field configuration space}
\label{sec:ccps}

In the body of the paper we considered an enlargement of the field
configuration space $\Gamma_0$ to a larger space $\Gamma_{\rm ext}$ which contains
additional unphysical metrics $(M,g_{ab}, \Omega)$ that are related to
metrics already in $\Gamma_0$ by diffeomorphisms and conformal
transformations.  This raises the question of what criterion
can one use to define field configuration spaces in general?  How much
gauge (here diffeomorphism and conformal) freedom can or should be
fixed?

A key consideration is that the phase space of the theory is
constructed from the field configuration space $\Gamma_0$ or
$\Gamma_{\rm ext}$ by modding out by degeneracy directions of the
presymplectic form \cite{LW,WZ,Harlow:2019yfa}.  The construction of the symmetry algebra also
mods out by these degeneracies (see \cref{footnote:trivial} above).
The degeneracy directions correspond to gauge transformations
(diffeomorphism or conformal) which vanish sufficiently rapidly near
the boundary.  Therefore, in defining the initial field
configuration space, it should not matter how much gauge freedom is
fixed, since any residual gauge freedom will be removed in the
construction of the final phase space and symmetry algebra.  However, one must be careful
that one fixes only ``true gauge'' degrees of freedom, that is,
degeneracy directions of the presymplectic form.

The question then is whether the standard configuration space
$\Gamma_0$ of Eq.\ (\ref{eq:fcs}) has already fixed some degrees of freedom
which are physical and not gauge (i.e. do not correspond to degeneracy
directions of the presymplectic form).  
Unfortunately, it is not straightforward to answer this question,
since as we have shown, for the relevant metric perturbations the presymplectic current
is either divergent on $\scri$, or if one uses the renormalized
presymplectic current of Ref.\ \cite{Cnew}, the presymplectic form may
or may not be covariant.
If we suppose for the sake of argument that it is covariant, then an
examination of Eq.\ (5.27) of Ref.\ \cite{Cnew} shows that the
presymplectic form on the extended phase space does not exhibit any
degeneracy directions.  This would argue in favor of the extended algebra.

%%===========================================================
\section{Discussion and Conclusions}
\label{conc}

Using the definition of asymptotically-flat spacetimes we showed how
the BMS algebra \(\mf b\) arises as the asymptotic symmetry algebra at
null infinity, emphasizing the role of the smoothness and topological
assumptions in the definition. A crucial role in our analysis was
played by the fact that the conformal class of metrics on a
\(2\)-sphere is unique up to diffeomorphisms. This can be used to show
that the conformal factor \(\Omega\) and the unphysical metric
\(g_{ab}\vert_\scri\) can be chosen to be universal in the class of
asymptotically-flat spacetimes, without loss of generality.

We then considered an extended class of spacetimes where \(g_{ab}\vert_\scri\) is not considered universal as proposed by Campiglia and Laddha \cite{CL}. In this class of spacetimes, the asymptotic symmetry algebra includes all smooth diffeomorphisms of a \(2\)-sphere. We showed, however, that the symplectic current of general relativity evaluated on such extended symmetries does not have a finite limit to null infinity, and that no local and covariant ambiguity in the choice of symplectic current cures this divergent behavior.  This suggests that the extension of the BMS symmetry algebra proposed by \cite{Campiglia:2014yka,CL} is ill-defined on the phase space at null infinity.  However, as discussed in the introduction, a possible loophole is the fact that imposing locality and covariance at all stages of the computation may be too strong a restriction, and instead one should only impose covariance on the symplectic form obtained by integrating the symplectic current.

%%======================================================
\section*{Acknowledgments}
This work is supported in part by the NSF grants PHY-1404105 and PHY-1707800 to Cornell University.

\appendix

%%=============================================================
\section{Metric on $\scri$ and conformal factor in a neighborhood can be chosen to be universal.}
%  Coordinate systems near \(\scri\)}
\label{sec:BS}

In this appendix, we prove the following property of
asymptotically flat spacetimes.
Suppose that we are given two different unphysical spacetimes $(M,
g_{ab},\Omega)$ and $(M', g_{ab}',\Omega')$ corresponding to two
different physical spacetimes.  By specializing to a neighborhood of
$\scri$ and using a diffeomorphism we can identify the background
manifolds.  In addition, by exploiting the diffeomorphism freedom $(g_{ab},\Omega)
\to (\psi_* g_{ab}, \psi_* \Omega)$ and the conformal
rescaling freedom $(g_{ab}, \Omega) \to (\omega^2 g_{ab}, \omega
\Omega)$, we can without loss of generality take
\be
\Omega =
\Omega'
\ee
in a neighborhood of $\scri$, as well as 
\be
\left. g_{ab} \right|_\scri = \left. g_{ab}' \right|_\scri 
\ee
together with the Bondi condition
\be
\left. \nabla_a \nabla_b \Omega \right|_\scri =0.
\ee
Thus, without loss of generality, we can take the conformal factor in
a neighborhood of $\scri$ and the
unphysical metric evaluated on $\scri$ to be universal, the same for all
asymptotically flat spacetimes.

The proof is based on constructing suitable coordinate systems in a
neighborhood of $\scri$.  
Let us pick \emph{any} asymptotically-flat physical spacetime \((\tilde M , \tilde g_{ab})\) and let \((M, g_{ab})\) be its conformal completion with some choice of conformal factor \(\Omega\) satisfying the Bondi condition (\ref{eq:Bondi-cond}). Pick \emph{any} cross-section \(S \cong \bb S^2\) of \(\scri\) and consider the induced metric \(q_{ab}\) at \(S\). Since the conformal class of metrics on \(\bb S^2\) is unique up to diffeomorphisms, there exists a unit round 2-metric $s_{ab}$  and a smooth positive function \(\varpi\) on \(S\) so that $q_{ab} = \varpi^2 s_{ab}$ on $S$.   We can now use the freedom (\ref{eq:conf-freedom}) in the choice of conformal factor at $S$ to make $\varpi=1$, so that $q_{ab} = s_{ab}$ on $S$.

Next, we choose coordinates $x^A = (x^1,x^2) = (\theta,\varphi)$ on $S$ so that this 2-metric $s_{AB}$ takes the standard form
\be
s_{AB} dx^A dx^B = d\theta^2 + \sin^2 \theta d\varphi^2.
\label{unitround}
\ee
We extend the coordinates \(x^A\) to all of \(\scri\) by imposing the condition
\be
n^a\nabla_a x^A \vert_\scri = 0.
\label{condt11}
\ee
We also define the function $u$ on $\scri$ by the conditions \(u\vert_S = 0\) and
\be
n^a\nabla_a u = 1.
\label{condt12}
\ee
This defines the coordinate system \((u,x^A)\) on \(\scri\). Note that on \(\scri\), \(\lie_n q_{ab} = 0\) (\cref{eq:q-cond}) and thus in our choice of conformal factor we have
that the induced metric takes the form (\ref{unitround})
on all of \(\scri\).  

We next define spacetime coordinates \((\Omega, u, x^A)\) in a neighborhood of $\scri$.
First, since \(\Omega\vert_\scri = 0 \) and \(\nabla_a \Omega\vert_\scri \neq 0\), we can use \(\Omega\) as a coordinate.
Next, we extend the coordinates \((u, x^A)\)
away from \(\scri\). There are many different choices of extension.
The extension we choose here leads to the familiar Bondi coordinates
in the physical spacetime (see also \cite{TW}). Consider a family of
null hypersurfaces transverse to \(\scri\) which intersect \(\scri\)
in the cross-sections \(S_u\) given by \(u = \text{constant}\). In a
sufficiently small neighborhood of \(\scri\), such null hypersurfaces
generate a null foliation. We first extend the coordinate \(u\) by
demanding that it be constant along  these null hypersurfaces. We
define
\be
l_a = \nabla_a u,
\ee
the null normal to these hypersurfaces,
which satisfies \(l^a l_a = 0\) and \(l^a n_a\vert_\scri =
1\) from the condition (\ref{condt12}). Then, we extend the angular
coordinates $x^A$ to a neighborhood of
\(\scri\) by demanding \(l^a \nabla_a x^A = 0\).

Finally, we specialize the definition of $\Omega$ off $\scri$ as follows.  
To extend
\(\Omega\) away from \(\scri\) we use the freedom in the conformal
factor away from \(\scri\) to demand that the \(2\)-spheres of
constant \(u\) and \(\Omega\) have the same area element as the unit sphere,
that is, if \(h_{AB}\) is the \(2\)-metric on the surfaces of constant
\(u\) and \(\Omega\) then we demand that \(\det h = \det s\) in the
\(x^A\)-coordinates. This fixes \(\Omega\) uniquely away from
\(\scri\). Thus we have set up a \emph{conformal Bondi} coordinate
system \((\Omega, u, \theta^A) \) in a neighborhood of \(\scri\) in
which the unphysical metric takes the form 
\be\label{eq:unphys-BS}
	ds^2 \equiv - W e^{2\beta} du^2 + 2 e^{2\beta} d\Omega du + h_{AB} (dx^A - U^A du )(dx^B - U^B du )
\ee
where \(W\), \(\beta\), \(h_{AB}\), and \(U^A\) are smooth functions of the coordinates \((\Omega, u, x^A)\). Note that the metric components \(g_{\Omega \Omega}\) and \(g_{\Omega A}\) vanish due to \(l^a l_a = l^a \nabla_a x^A = 0\). Now if we assume that the metric components in \cref{eq:unphys-BS} have an asymptotic expansion in integer powers of \(\Omega\) near \(\scri\), then from the construction of our coordinates, and using the condition (\ref{eq:nn-cond}) we have
\be\label{eq:metric-exp}\begin{aligned}
	W = O(\Omega^2) \eqsp \beta = O(\Omega) \eqsp U^A = O(\Omega), \\
    h_{AB} = s_{AB} + \Omega C_{AB} + O(\Omega^2) \eqsp s^{AB} C_{AB} = 0.
\end{aligned}\ee
If we further define \(r = \Omega^{-1}\) then we get the familiar
Bondi coordinate form\footnote{The usual Bondi coordinate expression incorporates two other conditions, 
  $W = \Omega^2 + O(\Omega^3)$, $U^A = O(\Omega^2)$, that are obtained by
  imposing the Einstein equations with the assumption (\ref{eq:st}) on the stress energy tensor.}
%More generally, if the two-metric $s_{AB}(x^C)$ is chosen to be other than the unit round 2-metric, then we have instead $W = \Omega^2 R^{(2)}/2 + O(\Omega^3)$, where $R^{(2)}$ is the two dimensional Ricci scalar of $s_{AB}$ \cite{Cnew}.}
for the physical metric \(\tilde g_{ab} =
\Omega^{-2} g_{ab} = r^2 g_{ab}\). Note that the asymptotic falloffs
in \(r\) used for the physical metric in Bondi coordinates follow from
\cref{eq:metric-exp} and are a direct consequence of asymptotic flatness and the construction of the coordinate system.\\

Since \(n^a\nabla_a u\vert_\scri = 1 \), the unphysical metric \(g_{ab}\) at \(\scri\) in these coordinates is
\be\label{eq:unphys-BS-scri}
    g_{ab}\vert_\scri \equiv 2d\Omega d u + s_{AB} dx^A dx^B.
\ee
We now identify \emph{all} asymptotically-flat spacetimes with each
other, in a neighborhood of \(\scri\), by identifying their points in
the coordinates \((\Omega, u, x^A)\) constructed above. With this
choice it follows that the conformal factor \(\Omega\) and the
unphysical metric (\ref{eq:unphys-BS-scri}) are
universal i.e. independent of the chosen asymptotically-flat physical
spacetime. Different asymptotically-flat physical metrics change only
the subleading metric components in \cref{eq:metric-exp} but agree to
leading order.\\ 

We emphasize that many of the choices made in constructing the Bondi
coordinates are irrelevant to this argument and are made just for
convenience. For instance, the choice of the unit-metric \(s_{ab}\) is
irrelevant. In \emph{any} asymptotically-flat spacetime, we can
instead choose the coordinates \(x^A\) and use the freedom \(\omega\)
in the conformal factor at \(\scri\) so that \(q_{ab} = q^{(0)}_{ab}\)
where \(q^{(0)}_{ab}\) is \emph{any fixed} metric on \(\bb S^2\). Then
we can proceed with the rest of the construction as before to conclude
that \(g_{ab}\vert_\scri\) is universal. The only important ingredient
is the fact that \(\bb S^2\) has a unique conformal class of
metrics up to diffeomorphisms. Similarly, the extension of the coordinates away from
\(\scri\) can also be chosen differently. For instance, consider the
null vector field \(l^a\vert_\scri \equiv \partial/\partial\Omega\)
transverse to the cross-sections \(S_u\) of \(\scri\). Instead of
choosing \(\Omega\) away from \(\scri\) to make the \(2\)-spheres have
unit area, we can extend this vector field away from \(\scri\) by demanding that \(l^a \equiv \partial/\partial\Omega\) be an affinely-parameterized null vector field i.e. \(l^al_a = 0\) and \(l^b \nabla_b l^a = 0\). Then we can extend the coordinates \((u, x^A)\) by parallel transport along \(l^a\). This defines a \emph{conformal Gaussian null} coordinate system in a neighborhood of \(\scri\) \cite{Hol-Th,HIW}. Now we can again identify all the asymptotically-flat spacetimes in these conformal Gaussian null coordinates to conclude that \(\Omega\) and \(g_{ab}\vert_\scri\) are universal in a neighborhood of \(\scri\). We can also develop an asymptotic expansion for the metric components in this coordinate system similar to \cref{eq:unphys-BS,eq:metric-exp} (see \cite{Hol-Th}) and again see that different asymptotically-flat physical metrics only change the subleading metric components but agree to leading order.\\

In conclusion, given the definition of asymptotic flatness and that the conformal class of metrics on \(\bb S^2\) is unique, it is always possible to identify all the unphysical spacetimes in a neighborhood of \(\scri\) so that \(\Omega\) and \(g_{ab}\vert_\scri\) are independent of the physical spacetime under consideration.

%%========================================================
\section{The extended BMS algebra does not contain any preferred translation subalgebra}
\label{sec:trans-ideal}

In this appendix we show that the extended BMS algebra does not
contain any preferred subalgebra (i.e.\ a Lie ideal) of
translations. Since the asymptotic symmetry algebra is common to all
spacetimes under consideration, its Lie bracket is independent of the
choice of background spacetime. Thus we can compute the Lie bracket on
\emph{any} choice of background spacetime, and in particular we can take the background physical spacetime to be Minkowski. Let us choose a conformal completion for Minkowski so that the induced metric \(q_{ab}\) on \(\scri\) is that of a unit-metric on \(\bb S^2\) and let \(\ms D\) denote the covariant derivative of \(q_{ab}\). Let \(u\) be an affine parameter along the null geodesics of \(n^a\) so that \(n^a\nabla_a u \vert_\scri = 1\).

From \cref{eq:bms-ext-cond}, any element \(\xi^a\) of the algebra \(\mf b_{\rm ext}\) can be written as
\be
    \xi^a = X^a + \tfrac{1}{2} (u-u_0) \ms D_b X^b n^a + f' n^a
\ee
where \(f'\) is any function on \(\bb S^2\) (representing a supertranslation), \(X^a\) is a vector field on \(\bb S^2\) while the function \(\alpha_{(\xi)} = \tfrac{1}{2} \ms D_a X^a\). The Lie bracket of a supertranslation \(f n^a \in \mf s\) and \(\xi^a\) is then
\be\label{eq:beta-defn}
    [f n, \xi]^a = \beta n^a \quad \text{where } \beta = - X^a \ms D_a f + \tfrac{1}{2} \ms D_a X^a f
\ee
It is straightforward to check that \(\lie_n \beta = 0\) and so \(\beta n^a\) is a supertranslation in \(\mf s\).

If translations are a Lie ideal in \(\mf b_{\rm ext}\) then \(\beta n^a\) would also be a translation whenever \(f n^a\) is translation. To investigate this we proceed as follows. Since \(f n^a\) is a translation, \(f\) is a \(\ell = 0,1\) spherical harmonic on \(\bb S^2\) --- it is well-known that the limit of translations in Minkowski spacetime to \(\scri\) are precisely such vector fields. Let \(X^a\) be a \(\ell'\)-vector harmonic so that
\be\label{eq:X-decomp}
    X^a = \ms D^a F + \varepsilon^{ab} \ms D_b G 
\ee
for some functions \(F\) and \(G\) which are \(\ell'\)-spherical harmonics. In the case \(\xi^a\) is an element of the BMS algebra \(\mf b\) so that \(X^a\) is an element of the Lorentz algebra \(\mf{so}(1,3)\), both \(F\) and \(G\) are spherical harmonics with \(\ell' = 1\). The function \(F\) corresponds to Lorentz boosts while \(G\) corresponds to Lorentz rotations. When \(\xi^a\) is an element of the extended BMS algebra \(\mf b_{\rm ext}\), \(\ell' \geq 1\) and then \(\ell' > 1 \) modes of \(F\) and \(G\) can be thought of as ``extended'' boosts and rotations.

Using the decomposition \cref{eq:X-decomp} in \cref{eq:beta-defn} we have
\be\begin{aligned}
    \beta & = - \ms D^a F \ms D_a f + \tfrac{1}{2} \ms D^2 F  f + \varepsilon^{ab} \ms D_a G \ms D_b f
\end{aligned}\ee
Now we wish to find the spherical harmonic mode \(L\) of \(\beta\) when \(f\) is a translation i.e. \(\ell = 0,1\)-harmonic mode while the harmonic mode of \(F\) and \(G\) can be \(\ell' \geq 1\). It is useful to consider the following different cases.
\paragraph*{Case 1: \(f\) is time translation, \(\ell = 0\)}
Then,
\be
    \beta = \tfrac{1}{2} \ms D^2 F f = - \tfrac{1}{2} \ell'(\ell'+1) F f
\ee
so
\be
    \ms D^2 \beta = - \ell' (\ell' + 1) \beta = - L(L+1)\beta
\ee
Thus, \(\beta = 0\) if \(F=0\) else \(\beta\) is a \(L = \ell'\) mode. When \(X^a \in \mf{so}(1,3)\), \(\ell' = 1\) and this can be interpreted as the fact that a time translation is invariant under Lorentz rotations given by \(G\) but changes by a spatial translation under Lorentz boosts given by \(F\).
 
\paragraph*{Case 2: \(f\) is spatial translation i.e. \(\ell =  1\), \(F=0\) and \(G \neq 0\)} 
Then we have,
\be
    \beta = \varepsilon^{ab} \ms D_a G \ms D_b f
\ee
and
\be\begin{aligned}
    \ms D^2 \beta & = \lb[ - \ell'(\ell'+1) - \ell(\ell+1) + 2 \rb] \beta + 2 \varepsilon^{ab} \ms D_c \ms D_a G \ms D^c \ms D_b f \\
    & = - \ell' (\ell' + 1) \beta = - L(L+1)\beta
\end{aligned}\ee
where in the last line we use \(\ell=1\) and that \(\ms D_a \ms D_b f = - q_{ab} f \) for such functions. Thus, \(\beta\) is a \(L = \ell'\) mode. Thus, when \(X^a \in \mf{so}(1,3)\), \(\ell' = 1\), a spatial translation changes by another spatial translation under Lorentz rotations given by \(G\).

\paragraph*{Case 3: \(f\) is spatial translation i.e \(\ell =  1\), \(F\neq 0\) and \(G=0\)} 
\be
    \beta = - \ms D_a F \ms D_a f + \tfrac{1}{2} \ms D^2 F f
\ee
To find the \(L\)-mode of \(\beta\), we multiply the above equation with the (complex conjugate) spherical harmonic \(\bar Y_{L,M}\) and integrate over \(\bb S^2\) to get (we have left the area element of the unit-metric on \(\bb S^2\) implicit for notational convenience)
\be
    \int \beta \bar Y_{L,M} = - \int \ms D^a F \ms D_a f \bar Y_{L,M} - \tfrac{1}{2} \ell' (\ell'+1) \int F f \bar Y_{L,M}
\ee
The first term on the right-hand-side can be rewritten using repeated integration-by-parts as
\be\begin{aligned}
    - \int \ms D^a F \ms D_a f \bar Y_{L,M}
    & = \int F \ms D^2 f \bar Y_{L,M} + \int F \ms D_a f \ms D^a \bar Y_{L,M} \\
    & = \int F \ms D^2 f \bar Y_{L,M} - \int  \ms D_a F f \ms D^a \bar Y_{L,M} - \int F f \ms D^2 \bar Y_{L,M} \\
    & = \int F \ms D^2 f \bar Y_{L,M} + \int  \ms D^2 F f \bar Y_{L,M} - \int F f \ms D^2 \bar Y_{L,M} + \int \ms D_a F \ms D^a f \bar Y_{L,M} \\
    - \int \ms D^a F \ms D_a f \bar Y_{L,M} 
     & = \tfrac{1}{2}\int F \ms D^2 f \bar Y_{L,M} + \tfrac{1}{2} \int  \ms D^2 F f \bar Y_{L,M} - \tfrac{1}{2} \int F f \ms D^2 \bar Y_{L,M} \\
    & = \tfrac{1}{2} \lb[ - \ell(\ell+1) - \ell'(\ell'+1) + L(L+1) \rb] \int F f \bar Y_{L,M}
\end{aligned}\ee
Thus, we have
\be
    \int \beta \bar Y_{L,M} = \lb[ - \tfrac{1}{2} \ell(\ell+1) - \ell'(\ell'+1) + \tfrac{1}{2} L(L+1) \rb] \int F f \bar Y_{L,M}
\ee
Expanding the functions \(F\) and \(f\) in terms of the corresponding spherical harmonics \(Y_{\ell', m'}\) and \(Y_{\ell,m}\) respectively, we can write the final integral in terms of the \(3\-j\)-symbols (see Sec.~34 \cite{DLMF}) (or in terms of the \emph{Clebsch-Gordon} coefficients, Sec.~3.7 \cite{Sakurai}) as
\be\begin{aligned}
    \int Y_{\ell',m'} Y_{\ell,m} \bar Y_{L,M} 
    & = (-1)^M \sqrt{ \frac{(2\ell +1) (2 \ell' + 1) (2L+1)}{4\pi}}
    \begin{pmatrix}
    \ell, & \ell', & L \\
    0, & 0, & 0 
    \end{pmatrix}
    \begin{pmatrix}
    \ell, & \ell', & L \\
    m, & m', & -M 
    \end{pmatrix}
\end{aligned}\ee
Since \(f\) is a spatial translation with \(\ell=1\), we have
\be
    \int \beta \bar Y_{L,M} \propto  \lb[ - 1 - \ell'(\ell'+1) + \tfrac{1}{2} L(L+1) \rb] 
    \begin{pmatrix}
    1, & \ell', & L \\
    0, & 0, & 0 
    \end{pmatrix}
    \begin{pmatrix}
    1, & \ell', & L \\
    m, & m', & -M 
    \end{pmatrix}
\ee
where we have ignored non-zero constant factors. The right-hand-side is non-vanishing if and only if (Sec.~34 \cite{DLMF})
\be\label{eq:L-constraints}\begin{aligned}
    - 1 - \ell'(\ell'+1) + \tfrac{1}{2} L(L+1) \neq 0 \\
    1 + \ell' + L \quad\text{is even} \\
    \ell' - 1 \leq L \leq \ell' + 1 \eqsp M = m + m'
\end{aligned}\ee
These conditions on \(L\) can be satisfied if and only if (we do not need the conditions on \(M\) for our argument)
\be
    L = 
    \begin{cases}
    0 & \text{for~} \ell' = 1 \\
    \ell' - 1 \text{~ or ~} \ell' + 1 & \text{for~} \ell' \geq 2 
    \end{cases}
\ee
Note that for the \(\ell' = 1\) case, the value \(L=\ell'+1 = 2\) is ruled out by the first condition in \cref{eq:L-constraints}. Thus, when \(X^a \in \mf{so}(1,3)\), \(\ell' = 1\), a spatial translation changes by a time translation under Lorentz boosts given by \(F\).\\

For the usual BMS algebra \(\mf b\) with \(X^a \in \mf {so}(1,3)\) and \(\ell' = 1\), we see that in each case \(\beta\) is a spherical harmonic with \(L = 0,1\) that is \(\beta n^a\) is a translation. Thus the translation subalgebra is preserved under the Lie bracket of \(\mf b\) i.e. there is a preferred \(4\)-dimensional Lie ideal of translations in \(\mf b\). For the extended BMS algebra \(\mf b_{\rm ext}\) with \(X^a \in \mf {diff}(\bb S^2)\) and \(\ell' \geq 2\), the translations \(f n^a\), in general, change by \(\beta n^a\) where \(\beta\) contains a spherical harmonic as high as \(L = \ell' +1\). Thus, translations are not preserved by the Lie bracket of \(\mf b_{\rm ext}\) and are not a preferred subalgebra (Lie ideal) of \(\mf b_{\rm ext}\). The above argument can be generalized to show that there is, in fact, no finite-dimensional Lie ideal of extended BMS algebra.

The absence of a preferred translation algebra poses a problem for the prescription used by \cite{CL} to define a symplectic form on \(\scri\). As discussed above, the boundary condition imposed by \cite{CL} near spatial infinity to obtain a finite symplectic form for \(\mf b_{\rm ext}\) is not invariant under general supertranslations, but is invariant under translations in a specific choice of Bondi coordinates. However, as we have shown, there is no preferred notion of pure translations in \(\mf b_{\rm ext}\). Thus, the translation invariance of the boundary condition in \cite{CL} is also unclear.

%%THE END

%\newpage
\bibliographystyle{JHEP}
\bibliography{extended-BMS}

\end{document}